\documentclass[a4paper,11pt]{article}
\usepackage{graphicx}
\usepackage{amsmath}
\usepackage{amssymb}
\usepackage{amsthm}
\usepackage{path}
\usepackage{enumerate}
\usepackage[pdfpagelabels, pagebackref, naturalnames]{hyperref}
\usepackage[font=small, format=hang, labelfont=bf]{caption}
\usepackage[subrefformat=parens, labelfont=default]{subfig}
\usepackage{a4wide}
\usepackage[linesnumbered,algoruled,vlined]{algorithm2e}

\title{{\bf Drawing (Complete) Binary Tanglegrams: \\
  {\Large Hardness, Approximation, Fixed-Parameter Tractability}}%
  \thanks{Work started at
    the 10th Korean Workshop on Computational Geometry, Dagstuhl,
    Germany, 2007. A preliminary version~\cite{bbbnosw-dcbth-09} of
    this paper was presented
    at the 16th International Symposium on Graph Drawing (GD'08).}}

\author{%
  Kevin~Buchin\thanks{Faculteit Wiskunde en Informatica, TU~Eindhoven,
    The~Netherlands. Email: \email{k.a.buchin@tue.nl}. K.~Buchin was
    supported by the Netherlands Organisation for Scientific Research
    (NWO) under project no.\ 639.022.707 and 642.065.503.}
  \and
  Maike~Buchin\thanks{Faculteit Wiskunde en Informatica, TU~Eindhoven,
    The~Netherlands. Email: \email{m.e.buchin@tue.nl}.  M.~Buchin was
    supported by the German Research Foundation (DFG) 
  under grant no.\ BU~2419/1-1 and by the Netherlands Organisation for
  Scientific Research (NWO) under project no.\ 642.065.503.}
  \and
  Jaroslaw~Byrka\thanks{Institute of Computer Science, University of
    Wroclaw, Poland.  Email: \email{jby@ii.uni.wroc.pl}.  J.~Byrka was
    partially supported by MNiSW grant number N~N206 368839,
    2010-2013. His research was partially conducted at TU Eindhoven
    and at EPFL, Lausanne.}
  \and
  Martin~N\"ollenburg\thanks{Institute of Theoretical Informatics,
    Karlsruhe Institute of Technology (KIT), Germany. Email:
    \email{noellenburg@kit.edu}.  M.~N\"ollenburg was
    supported by grant WO~758/4-3 of the German Research Foundation
    (DFG).}
  \and
  Yoshio~Okamoto\thanks{Graduate School of Information Science and
    Engineering, Tokyo Institute of Technology, Japan.  Email:
    \email{okamoto@is.titech.ac.jp}.  Y.~Okamoto was partially
    supported by Grant-in-Aid for Scientific Research and Global COE
    Program ``Computationism as a Foundation for the Sciences'' from
    Ministry of Education, Science and Culture, Japan, and Japan
    Society for the Promotion of Science.}
  \and
  Rodrigo~I.~Silveira\thanks{Dept. Matem\`{a}tica Aplicada II,
    Universitat Polit\`{e}cnica de Catalunya, Spain.  Email:
    \email{rodrigo.silveira@} \email{upc.edu}.  R.~Silveira was supported by
    the Netherlands Organisation for Scientific Research (NWO).}
  \and
  Alexander~Wolff\thanks{Lehrstuhl~I, Institut f\"ur Informatik,
    Universit\"at W\"urzburg, Germany.  WWW:
    \email{www1.informatik.uni-} 
    \email{wuerzburg.de/en/staff/wolff\_alexander}}
}

\date{} 

\theoremstyle{plain}
\newtheorem{definition}{Definition}[section]
\newtheorem{theorem}[definition]{Theorem} 
\newtheorem{lemma}[definition]{Lemma}
\newtheorem*{observation}{Observation}

\newcommand{\true}{\emph{true}}
\newcommand{\false}{\emph{false}}
\newcommand{\ttree}[1]{\ensuremath{\langle #1 \rangle}}
\newcommand{\swapl}{\ensuremath{\text{swp}_{S}}}
\newcommand{\swapr}{\ensuremath{\text{swp}_{T}}}

\DeclareMathOperator{\cross}{cr}
\DeclareMathOperator{\cl}{cl}
\DeclareMathOperator{\lca}{lca}

\newenvironment{pf}{\begin{proof}}{\end{proof}}

\graphicspath{{fig-arxiv/}}

\newenvironment{keywords}{\vspace*{2ex}\noindent\small\textbf{Keywords.}\hspace{1.5ex}}{\vspace*{3ex}\normalsize}

\renewcommand{\figurename}{Fig.}
\newcommand{\email}[1]{\texttt{#1}}

\def\comment#1{}

\def\withcomments{
   \newcounter{mycommentcounter}
   \def\comment##1{\refstepcounter{mycommentcounter}%
    \ifhmode%
     \unskip%
     {\dimen1=\baselineskip \divide\dimen1 by 2 %
       \raise\dimen1\llap{\tiny -\themycommentcounter-}}\fi%
     \marginpar{\renewcommand{\baselinestretch}{0.8}%
       \footnotesize [\themycommentcounter]: \raggedright ##1}}
   }
\withcomments

\renewcommand*{\backref}[1]{}
\renewcommand*{\backrefalt}[4]{%
  \small
  \ifcase #1 %
  [not cited]%
  \or
  [p.~#2]%
  \else
  [pp.~#2]
  \fi
}

\hypersetup{%
  pdftitle={}, 
  pdfauthor={Alexander Wolff}, 
  pdfborder={0 0 1},
  pdfcreator={}, pdfproducer={},
  citebordercolor={0 .667 0},          
  linkbordercolor={.5812 .0665 .0659}, 
  urlbordercolor={0 0 .667}            
}

\begin{document}

\maketitle

\begin{abstract}
  A \emph{binary tanglegram} is a drawing of a pair of
  rooted binary trees
  whose leaf sets are in one-to-one correspondence; matching leaves
  are connected by inter-tree edges.
  For applications, for example, in phylogenetics, it is essential
  that both trees are drawn
  without edge crossings and that the inter-tree edges
  have as few crossings as possible.  It is known that finding a
  tanglegram with the minimum number of crossings is NP-hard and that the
  problem is fixed-parameter tractable with respect to that
  number.

  We prove that under the Unique Games Conjecture there is no
  constant-factor approximation for binary trees.  We show
  that the problem is NP-hard even if both trees are complete binary
  trees.  For this case we give an $O(n^3)$-time 2-approximation and a
  new, simple fixed-parameter algorithm.  
  We show that the maximization version of the dual problem for
   binary trees can be reduced to a version of \textsc{MaxCut}
  for which the algorithm of Goemans and Williamson yields a
  $0.878$-approximation.
\end{abstract}


\newpage

\section{Introduction}

\begin{figure}[htb]
  \subfloat[arbitrary layout\label{sfg:permuted}]%
  {\includegraphics[width=.48\textwidth]{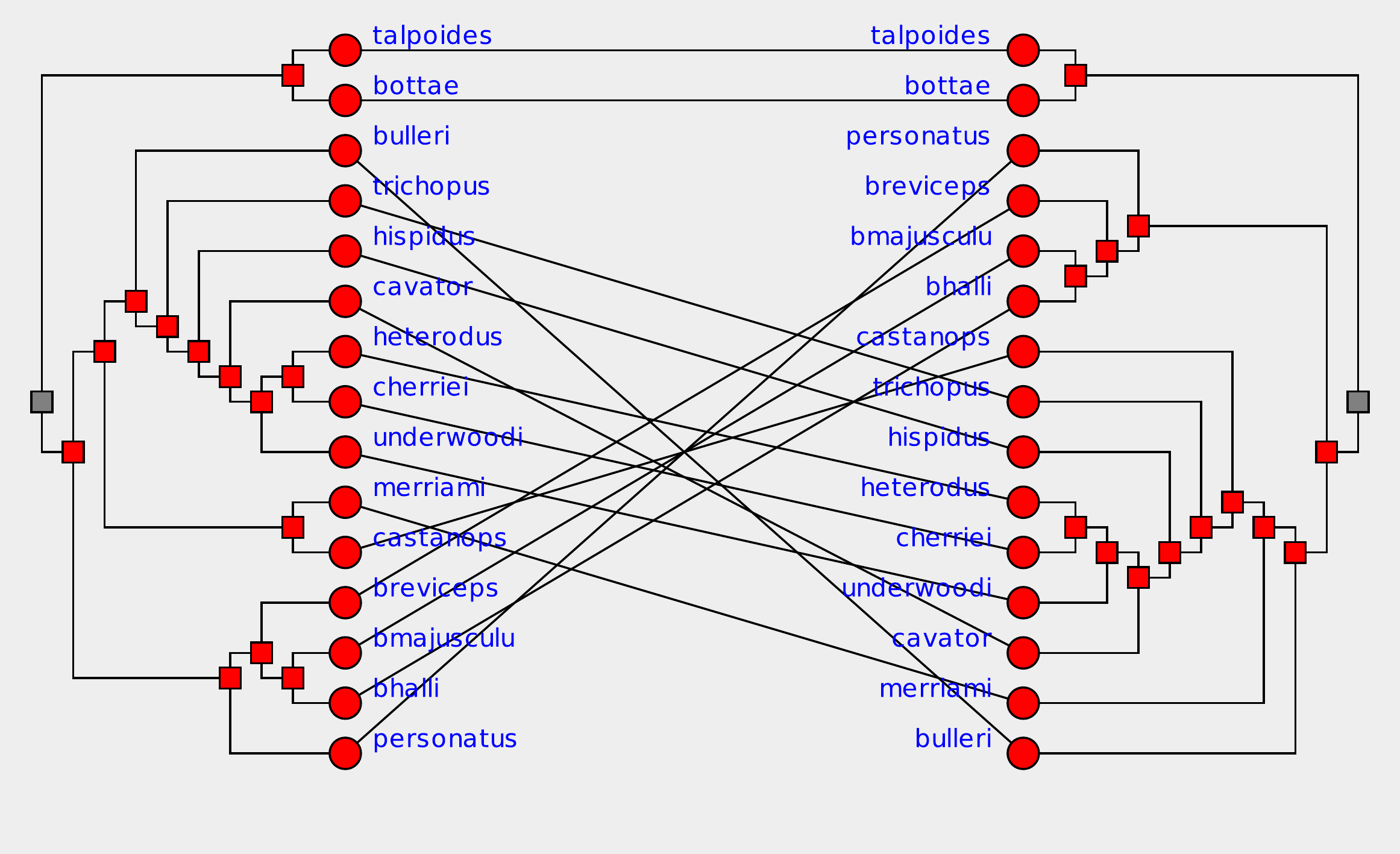}}
  \hfill
  \subfloat[layout by our 2-approximation algorithm\label{sfg:untangled}]%
  {\includegraphics[width=.48\textwidth]{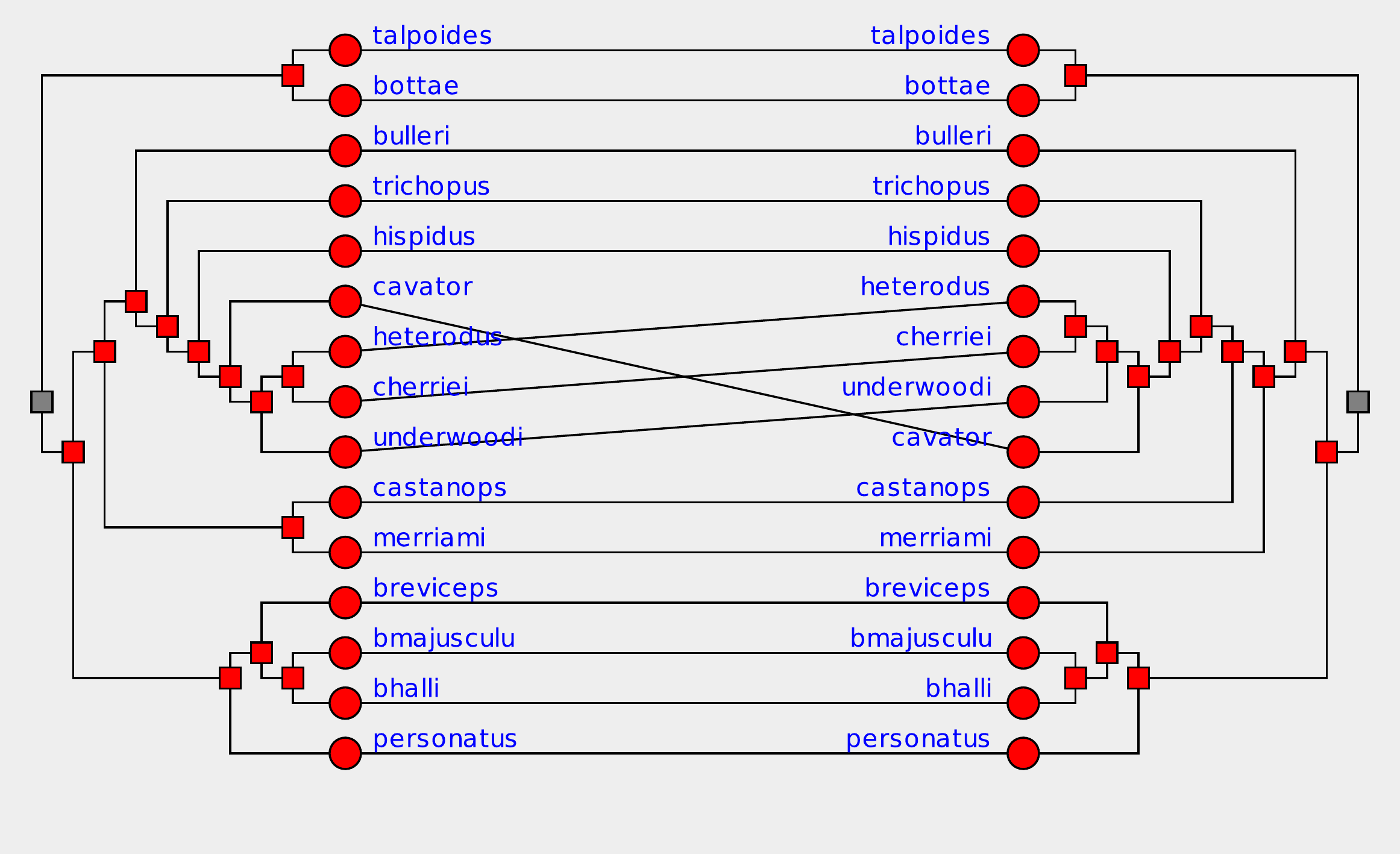}}
  \caption{A binary tanglegram showing two evolutionary
    trees for lice of pocket gophers~\cite{hsvsdn-drmec-94}.}
  \label{fig:example}
\end{figure}

In this paper we are interested in drawing so-called
\emph{tanglegrams}~\cite{p-ttpcc-02}, that is, comparative drawings of
pairs of rooted trees
whose leaf sets are in one-to-one correspondence.  The need to
visually compare pairs of trees arises in applications such as the
analysis of software projects, phylogenetics, or clustering.  In the
first application, trees may represent package-class-method
hierarchies or the decomposition of a project into layers, units, and
modules~\cite{hw-vchod-08}.  The aim is to analyze changes in
hierarchy over time or to compare human-made decompositions with
automatically generated ones.  Whereas trees in software analysis can
have nodes of arbitrary degree, trees from our second application,
that is, (rooted) phylogenetic trees, are binary trees. This makes
binary tanglegrams an interesting special case, see
\figurename~\ref{fig:example}. Tanglegrams in phylogenetics are used,
for example, to study cospeciation~\cite{p-ttpcc-02} or to compare
evolutionary trees for the speciation of a single lineage but from
different tree building methods.  Hierarchical clusterings,
our third application, are usually visualized by a binary tree-like
structure called \emph{dendrogram}, where elements are represented by the
leaves and each internal node of the tree represents the cluster
containing the leaves in its subtree. Pairs of dendrograms stemming
from different clustering processes of the same data can be compared
visually using tanglegrams.
Note that we are interested in minimizing the number of crossings for
visualization purposes.  The minimum, as a number, is not primarily
intended to be a tree-distance measure (since, for example, a crossing
number of zero does not mean that two trees are equal).  Examples of 
such measures are nearest-neighbor interchange and 
subtree transfer \cite{dhjltz-dbpt-97}.

Let $S$ and $T$ be two rooted, unordered,
$n$-leaf trees with node sets $V(S)$ and $V(T)$, edge sets $E(S)$ and $E(T)$,
and leaf sets $L(S) \subseteq V(S)$ and $L(T) \subseteq V(T)$,
respectively. In the remainder of the paper, unless
explicitly stated otherwise, trees are considered to be rooted and unordered. We say
that the pair of trees $\ttree{S,T}$ is \emph{uniquely leaf-labeled} if there are two
bijective labeling functions $\lambda_S \colon L(S) \rightarrow \Lambda$
and $\lambda_T \colon L(T) \rightarrow \Lambda$, where $\Lambda = \{1,
\ldots, n\}$ is a set of labels. For a uniquely leaf-labeled pair of trees
$\ttree{S,T}$ we define the set $E(S,T) = \{ uv \mid u \in L(S),\, v
\in L(T),\, \lambda_S(u) = \lambda_T(v)\}$ of \emph{inter-tree edges},
where each edge in $E(S,T)$ connects two leaves with the same label.

\smallskip
\noindent\textbf{Tanglegram Layout Problem\footnote{The name follows the common
    terminology in the biology
    literature~\cite{p-ttpcc-02,lprvz-stapt-07,vajg-utctd-10}. Note
    that the problem has also been called the two-tree crossing
    minimization problem~\cite{fkp-ctvcm-05} or the stratified tree
    ordering problem~\cite{ds-oloth-04}.} (TL)}
  Given a uniquely leaf-labeled pair of trees $\ttree{S,T}$, find a
  \emph{tanglegram} of $\ttree{S,T}$, that is, a drawing of the graph
  $G=(V(S)\cup V(T), E(S) \cup E(T) \cup E(S,T))$ in the plane, with
  the following properties:
  \begin{enumerate}
  \item The subdrawing of $S$ is a plane, \emph{leftward} drawing of $S$ with
    the leaves $L(S)$ on the line $x=0$ and each parent node strictly
    to the left of all its children;
  \item the subdrawing of $T$ is a plane, \emph{rightward} drawing of $T$ with
    the leaves $L(T)$ on the line $x=1$ and each parent node strictly
    to the right of all its children;
  \item the inter-tree edges $E(S,T)$ are drawn as straight-line segments;
  \item the number of crossings (between inter-tree edges) in the
    drawing is minimum.
  \end{enumerate}
\smallskip

In this paper we consider binary tanglegrams,
that is, tanglegrams that consist of two rooted binary trees.  We call the
restriction of TL to binary trees the \emph{binary TL problem}.
We say that a rooted binary tree is \emph{complete} (or \emph{perfect})
if all its leaves have the same distance to the root.  Accordingly, we
call the restriction of the binary TL problem to complete binary trees
the \emph{complete binary TL problem}. Figure~\ref{fig:example} shows two binary tanglegrams
for the same pair of trees, an arbitrary tanglegram and one with a
minimum number of crossings.

The TL problem is purely combinatorial:
Given a tree~$T$, we say that a linear order of
$L(T)$ is \emph{compatible} with~$T$ if for each node~$v$ of~$T$ the
nodes in the subtree of~$v$ form an interval in the order. For a
binary tree~$T$ the linear
orders of~$L(T)$ that are compatible with~$T$ are exactly those orders
that can be obtained from an initial plane leftward (or rightward)
drawing of $T$ by performing a sequence of subtree \emph{swaps} that flip the
order of the two child subtrees at an internal node.
Given a permutation~$\pi$ of $\{1,\dots,n\}$, we call~$(i,j)$ an
\emph{inversion} in $\pi$ if $i<j$ and $\pi(i)>\pi(j)$. For fixed
orders $\sigma$ of~$L(S)$ and $\tau$ of~$L(T)$ we define the
permutation~$\pi_{\tau,\sigma}$, which for a given position in~$\tau$
returns the position in~$\sigma$ of the leaf having the same label.  Now 
the TL problem consists in finding an order~$\sigma$ of~$L(S)$ compatible
with~$S$ and an order~$\tau$ of~$L(T)$ compatible with~$T$ such that
the number of inversions in~$\pi_{\tau,\sigma}$ is minimum.

\paragraph{Related problems.} In graph drawing the so-called
\emph{two-sided crossing minimization problem} (2SCM) is an important
problem that occurs when computing layered graph layouts.  Such
layouts were introduced by Sugiyama et al.~\cite{stt-mvuhs-81}
and are widely used for drawing hierarchical graphs. In 2SCM,
vertices of a bipartite graph are to be placed on two parallel lines
(called \emph{layers}) such that vertices on one line are adjacent
only to vertices on the other line.  As in TL the objective is to
minimize the number of edge crossings provided that edges are drawn as
straight-line segments.  In one-sided crossing minimization (1SCM) the
order of the vertices on one of the layers is fixed.  Even 1SCM is
NP-hard~\cite{ew-ecdbg-94}.
In contrast to TL, a vertex in an instance of 1SCM or 2SCM can have
several incident edges and the linear order of the vertices in the
non-fixed layer is not required to be compatible with a tree.
The following is known about 1SCM.
The median heuristic of Eades and
Wormald~\cite{ew-ecdbg-94} yields a 3-approximation and a randomized
algorithm of Nagamochi~\cite{n-ib1sm-05} yields an expected
1.4664-approximation.
Dujmovi{\v c} et al.~\cite{dfk-fpa1s-08} give an FPT 
algorithm that runs in $O^\star(1.4664^k)$ time, where $k$ is
the minimum number of crossings in any 2-layer drawing of the given
graph that respects the vertex order of the fixed layer.  The
$O^\star(\cdot)$-notation ignores polynomial factors.

\paragraph{Previous work.}
Dwyer and Schreiber~\cite{ds-oloth-04} draw series of related
tanglegrams in 2.5 dimensions.
Each tree is drawn on a plane, and the planes are stacked on top of each other.
They consider a one-sided version of
binary TL by fixing the layout of the first tree in the stack,
and then, plane-by-plane, computing the leaf order of the next tree in
$O(n^2 \log n)$ time each.
Binary TL is also studied by Fernau et al.~\cite{fkp-ctvcm-05},
although they refer to it as the \emph{two-tree crossing minimization}
problem.  They show that binary TL is NP-hard and give a
fixed-parameter algorithm that runs in
$O^\star(c^k)$ time, where~$c$ is a constant estimated to be~$1024$ 
and $k$ is the minimum number of crossings in any drawing of
the given tanglegram.
In addition, they show that the one-sided version of binary TL can
be solved in $O(n \log^2 n)$ time.  This improves on the result of
Dwyer and Schreiber~\cite{ds-oloth-04}.  Fernau et al.\
also make the simple observation that the edges of the tanglegram can
be directed from one root to the other.  Thus the existence of a
crossing-free tanglegram can be verified using a linear-time upward-planarity
test for single-source directed acyclic graphs~\cite{bdmt-dupts-98}.
Later, apparently not being aware of the above mentioned results, Lozano et
al.~\cite{lprvz-stapt-07} give a quadratic-time algorithm for the same
special case, to which they refer as \emph{planar tanglegram layout}.
Holten and van Wijk~\cite{hw-vchod-08} present a visualization tool
for general tanglegrams that heuristically reduces crossings (using
the barycenter method for 1SCM on a per-level base) and draws
inter-tree edges in bundles (using B{\'e}zier curves).

\paragraph{Our results.}
We first analyze
the complexity of binary TL, see Section~\ref{sec:complexity}.
We show that binary TL is essentially as hard as the
\textsc{MinUncut} problem.  If the (widely accepted) Unique Games
Conjecture holds, it is NP-hard to approximate
\textsc{MinUncut}---and thus binary TL---within any
constant factor~\cite{kv-ugcig-05}.
This motivates us to consider \emph{complete} binary TL.  It turns out
that this special case has a rich structure.  We start our
investigation by giving a new reduction from \textsc{Max2Sat} that
establishes the NP-hardness of complete binary TL.

The main result of this paper is a simple recursive factor-2
approximation algorithm for complete binary TL, see
Section~\ref{sec:approximation}.  It runs in $O(n^3)$ time and extends
to $d$-ary trees.
Our algorithm can also process non-complete binary tanglegrams---without
guaranteeing any approximation ratio.  It works 
well in practice
and is quite fast when combined with branch-and-bound~\cite{nvwh-dbtee-09}.

Next we consider a dual problem: maximize the number of edge pairs
that do \emph{not} cross.  We show that this problem (for binary
trees) can be reduced to a version of
\textsc{MaxCut} for which the algorithm of Goemans and Williamson~\cite{gw-iaamc-95}
yields a $0.878$-approximation.

Finally, we investigate the parameterized complexity of complete
binary TL.  Our parameter is the number~$k$ of crossings in an optimal
drawing.  We give a new FPT algorithm for complete binary TL that is
much simpler and faster than the FPT
algorithm for binary TL by Fernau et
al.~\cite{fkp-ctvcm-05}.  The running time of our algorithm is $O(4^k
n^2)$, see Section~\ref{sec:fpt}.
An interesting feature of the algorithm is that the parameter does
\emph{not} drop in each level of the recursion.

\paragraph{Subsequent work.} %
Since the presentation of the preliminary
version~\cite{bbbnosw-dcbth-09} of this
work, the TL problem has received a lot of
attention.  We briefly summarize these recent developments.
B\"{o}cker et al.~\cite{bhtw-ffpa-09} present a fixed-parameter
algorithm for binary TL that runs in $O(2^k n^4)$ time.  They further
give a kernel-like bound for complete binary TL.  Baumann et
al.~\cite{bbl-ebcmt-10} study a generalized version of TL, in which
the leaves no longer have to be in one-to-one correspondence; instead,
the inter-tree edges may form any bipartite graph.  They show how to
formulate the problem as a quadratic linear-ordering problem with
additional side constraints.  Bansal et al.~\cite{bcef-gbt-09} study
the same generalization, but restricted to binary TL.  For the
one-sided case (where the leaf order of one tree is fixed), they give
a polynomial-time algorithm.  On instances of (non-generalized)
one-sided binary TL, their algorithm runs in $O(n \log^2 n/\log \log
n)$ time, improving on the algorithm of Fernau et al.  Finally,
Venkatachalam et al.~\cite{vajg-utctd-10} give an $O(n \log n)$-time
solution for the same problem.

\section{Complexity} \label{sec:complexity}

In this section we consider the complexity of binary TL, which Fernau
et al.~\cite{fkp-ctvcm-05} have shown to be NP-complete.  We
strengthen their findings in two ways.  First, we show that it is
unlikely that an efficient constant-factor approximation for
binary TL exists.  Second, we show that TL remains hard even when
restricted to \emph{complete} binary tanglegrams.

We start by showing that binary TL is essentially as hard as
\textsc{MinUncut}, the dual formulation of the classic
\textsc{MaxCut} problem~\cite{gj-ci-79}.
This result relates the existence of a
constant-factor approximation for binary TL to the Unique Games Conjecture
(UGC). The UGC was introduced by Khot~\cite{k-pu2p1-02} in the context
of interactive proofs. It concerns a scenario with two provers and a
single round of answers to a question of the verifier.  The word
``unique'' refers to the strategy of the verifier, who for any fixed
answer of one of the provers will accept the proof only if the other
prover gives the unique second part of the proof. The provers cannot
communicate with each other.  Still they want to maximize the
probability of the proof being accepted given that questions of the
verifier are drawn randomly from a given distribution. The UGC
states that it is NP-hard to decide whether the optimal strategy
of the provers gives them a high probability of success.

The UGC became famous when it was discovered that it implies optimal
hardness-of-approxi\-ma\-tion results for problems such as \textsc{MaxCut}
and \textsc{VertexCover}, and forbids constant factor-approxi\-ma\-tion
algorithms for problems such as \textsc{MinUncut} and
\textsc{SparsestCut}~\cite{kv-ugcig-05}.
We reduce the \textsc{MinUncut} problem to the binary TL problem, which, by the
result of Khot and Vishnoi~\cite{kv-ugcig-05}, makes it unlikely that
an efficient constant-factor approximation for binary TL exists.

The \textsc{MinUncut} problem is defined as follows.  Given an
undirected graph $G=(V,E)$, find a partition $(V_1,V_2)$ of the
vertex set $V$ that minimizes the number of edges that are not cut by
the partition, that is, $\min_{(V_1,V_2)} |\{uv \in E : \{u,v\} \subseteq
V_1 \text{ or } \{u,v\} \subseteq V_2\}|$. Note that an optimal
solution for \textsc{MinUncut} of a graph $G$ is at the same time an optimal
solution for \textsc{MaxCut} of $G$. Nevertheless, the \textsc{MinUncut}
problem is more difficult to approximate.

\begin{theorem}
  \label{thm:hard}
  Under the Unique Games Conjecture it is NP-hard to approximate the
  TL problem for binary trees within any constant factor.
\end{theorem}

\begin{pf}
  As mentioned above, we reduce from the \textsc{MinUncut} problem.
  Our reduction is similar to the reduction in the NP-hardness proof
  by Fernau et al.~\cite{fkp-ctvcm-05}.

  Consider an instance $G=(V,E)$ of the \textsc{MinUncut} problem.  We
  construct a binary TL instance $\ttree{S,T}$ as follows.  The two
  trees $S$ and $T$ are isomorphic and there are three groups of edges
  connecting leaves of $S$ to leaves of $T$. For simplicity of
  exposition, we permit multiple inter-tree edges between a pair of
  leaves and also an inter-tree connection of a leaf to many other
  leafs in the other tree. In the actual trees, we replace each such
  meta-leaf by a binary tree with the appropriate number of regular
  leaves.

  Let $V = \{v_1,v_2,\ldots,v_n\}$ be the vertex set of the graph~$G$
  that constitutes our \textsc{MinUncut} instance.  Then we construct
  both~$S$ and~$T$ as follows.  We start with what we call the
  \emph{backbone} path 
  $\langle v_{11},v_{12},v_{21},v_{22},\ldots, v_{n1},v_{n2},a
  \rangle$ from the root node~$v_{11}$ to a central leaf~$a$.
  Additionally, for $i \in \{1,\ldots ,n\}$ and $j \in \{1,2\}$, we
  attach each node~$v_{ij}$ to a leaf~$\ell_{ij}$.  (The construction
  of~$S$ and~$T$ is illustrated, for the complete graph
  $K_3=(\{v_1,v_2,v_3\},\{v_1v_2,v_2v_3,v_3v_1\})$, in
  \figurename~\ref{fig:apx-hard}.)  In the remainder of this proof,
  where needed, we use a superscript to denote the tree to which a
  leaf belongs.  The inter-tree edges between~$S$ and~$T$ form the
  following three groups.
  \begin{itemize}
  \item Group A contains $n^{11}$ edges connecting the central leaves
    of the two trees. 
  \item Group B contains, for each $v_i \in V$, $n^7$ edges connecting
    $\ell^S_{i1}$ with $\ell^T_{i2}$ and $n^7$ edges connecting
    $\ell^S_{i2}$ with $\ell^T_{i1}$.
  \item Group C contains, for each $v_iv_j \in E$, a single edge from
    $\ell^S_{i1}$ to $\ell^T_{j1}$. 
  \end{itemize}
  Note that group~C contains possibly more than one inter-tree edge
  attached to a single leaf in the described tree. The actual, final
  tree is then obtained by replacing each leaf of the tree described
  above by a tree with $O(n)$ new leaves such that no two inter-tree
  edges share a leaf.  This replacement may cause new crossings, but
  no more than $O(n^2)$.  Hence, these crossings can be neglected in
  the analysis, where only terms of order~$n^{11}$ will matter.

  \begin{figure}[tb]
    \centering
    \includegraphics[width=\textwidth]{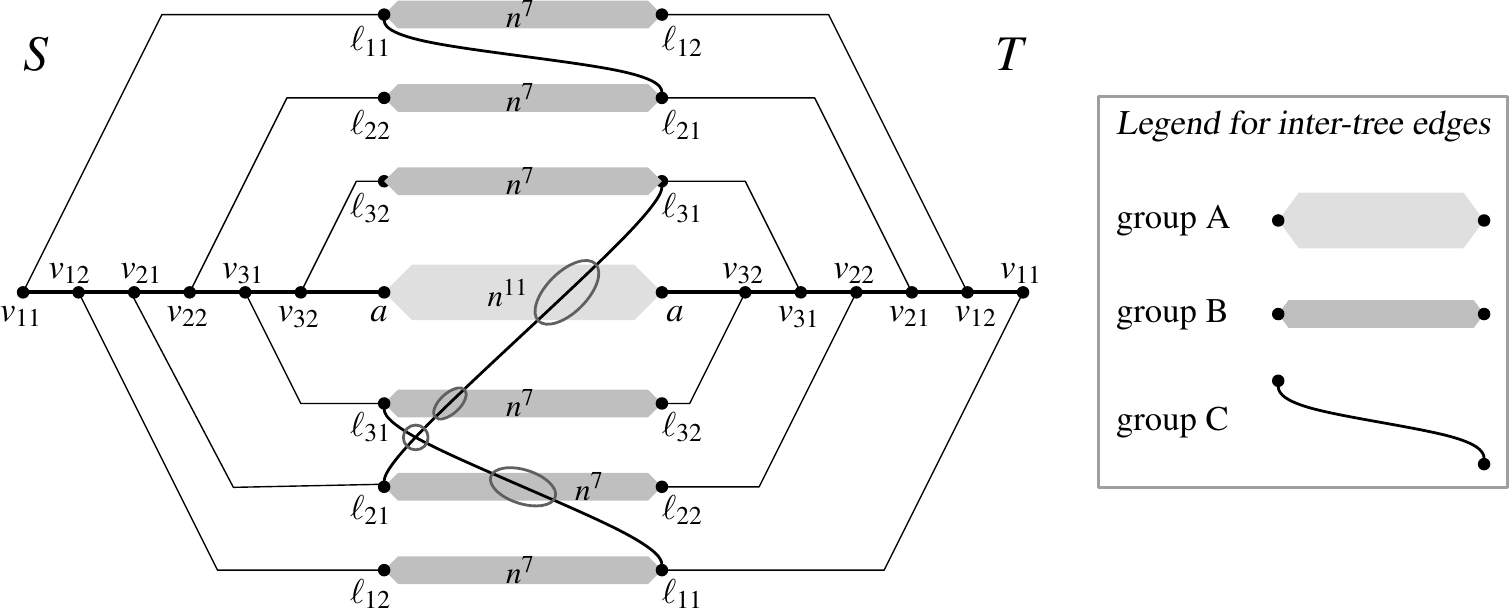}
    \caption{Binary TL instance corresponding to the graph $K_3$ and the cut
      $(\{v_1\},\{v_2,v_3\})$.  The crossings of the inter-tree edges
      are marked by gray ellipses.}
    \label{fig:apx-hard}
  \end{figure}

  Next, we show how to transform any partition in $G$ into a solution
  of the corresponding binary TL instance $\ttree{S,T}$. For our
  reduction we will apply this transformation to the partition of an
  optimal solution to the given \textsc{MinUncut} instance. Let
  $(V_1^*,V_2^*)$ be the given partition of $G$ and suppose that $k$
  is the number of edges that are not cut.  We now construct a drawing
  of $\ttree{S,T}$ such that at most $k \cdot n^{11} + O(n^{10})$
  pairs of edges cross. (In the example of
  \figurename~\ref{fig:apx-hard} we consider the cut
  $(\{v_1\},\{v_2,v_3\})$ with the uncut edge $v_2v_3$.)  We simply
  draw, for each vertex $v_i \in V_1^*$, the leaves~$\ell^S_{i1}$
  and~$\ell^T_{i2}$ above the backbones, and the leaves~$\ell^S_{i2}$
  and~$\ell^T_{i1}$ below the backbones.  Symmetrically, for each vertex
  $v_i \in V_2^*$, we draw the leaves~$\ell^S_{i1}$ and~$\ell^T_{i2}$
  below the backbones, and the leaves~$\ell^S_{i2}$ and~$\ell^T_{i1}$
  above the backbones.  Let us check the resulting number of
  crossings.  There are $k \cdot n^{11}$ A--C crossings, no A--B
  crossings, at most $|E| \cdot n^8 \in O(n^{10})$ B--C crossings, and
  at most $|E|^2 \in O(n^4)$ C--C crossings. (In
  \figurename~\ref{fig:apx-hard}, we have $k=1$, $|E|=3$, and $n^{11}
  + 2n^7 + 1$ crossings in total.)

  Now, suppose there exists, for some constant~$\alpha$, an
  $\alpha$-approximation algorithm for the binary TL problem.
  Applying this algorithm to the instance $\ttree{S,T}$ defined above
  yields a drawing $D(S,T)$ with at most $\alpha \cdot k \cdot n^{11}
  + O(n^{10})$ crossings.  Let us assume that $n$ is much larger
  than~$\alpha$ and than any of the constants hidden in the
  $O(\cdot)$-notation.  We show that from such a drawing $D(S,T)$ we
  would be able to reconstruct a cut $(V_1,V_2)$ in $G$ with at most
  $\alpha \cdot k$ uncut edges.  First, observe that
  nodes~$\ell^S_{i1}$ and~$\ell^T_{i2}$ must be drawn either both
  above or both below the backbones, otherwise there would be $n^{18}$
  A--B crossings. Similarly, $\ell^S_{i2}$ must be on the same side
  as~$\ell^T_{i1}$.  Next, observe that nodes~$\ell^S_{i1}$
  and~$\ell^S_{i2}$ must be drawn on different sides of the backbones,
  otherwise there would be $O(n^{14})$ B--B crossings.  Finally,
  observe that if we interpret the set of vertices $v_i$ for which
  $\ell^S_{i1}$ is drawn above the backbone as the set $V_1$ of a
  partition of $G$ and its complement as the set $V_2$, then this
  partition leaves at most $\alpha \cdot k$ edges from $E$ uncut.

  Hence, an $\alpha$-approximation for the binary TL problem would
  provide an $\alpha$-approximation for the \textsc{MinUncut} problem,
  which would contradict the UGC.
\end{pf}

The above negative result for binary TL is our motivation to
investigate the complexity of complete binary TL.  It turns out that
even this special case is hard.  Unlike Fernau et
al.~\cite{fkp-ctvcm-05}, who showed hardness of binary TL by a reduction
from \textsc{MaxCut} using extremely unbalanced trees, we use a quite
different reduction from a variant of \textsc{Max2Sat}.

\begin{theorem}
  The TL problem is NP-complete even for complete binary trees.
\end{theorem}
\begin{pf}
  Recall the \textsc{Max2Sat} problem which is defined as
  follows. Given a set $U = \{x_1, \ldots, x_n\}$ of Boolean
  variables, a set $C = \{c_1, \ldots, c_m\}$ of disjunctive clauses
  containing two literals each, and an integer $K$, the question is
  whether there is a truth assignment of the variables such that at
  least $K$ clauses are satisfied.  We consider a restricted version
  of \textsc{Max2Sat}, where each variable appears in at most three
  clauses.  This version remains NP-complete~\cite{rrr-snpcm-98}.

  Our reduction constructs two complete binary trees $S$ and $T$, in
  which certain aligned subtrees serve as variable gadgets and others
  as clause gadgets. We further determine an integer~$K'$ such that
  the instance~$\ttree{S,T}$ has less than~$K'$ crossings if
  and only if the corresponding \textsc{Max2Sat} instance has a truth
  assignment that satisfies at least $K$ clauses.

  The high-level structure of the two trees is depicted in
  \figurename~\ref{fig:bintrees-highlevel}.  From top to bottom, the four
  subtrees at level~2 on both sides are a clause subtree, a variable
  subtree, another clause subtree, and finally a dummy subtree.  The
  subtrees are connected to each other by inter-tree edges such that in any
  optimal solution they must be aligned in the depicted (or mirrored)
  order.  Each clause gadget appears twice, once in each clause
  subtree, and is connected to the variable gadgets belonging to its
  two literals.  Pairs of corresponding gadgets in $S$ and $T$ are
  connected to each other.  Finally, non-crossing dummy edges connect
  unused leaves in order to make $S$ and $T$ complete.  In the
  following, we describe the gadgets in more detail.

  \begin{figure}[b]
    \centering
    \includegraphics{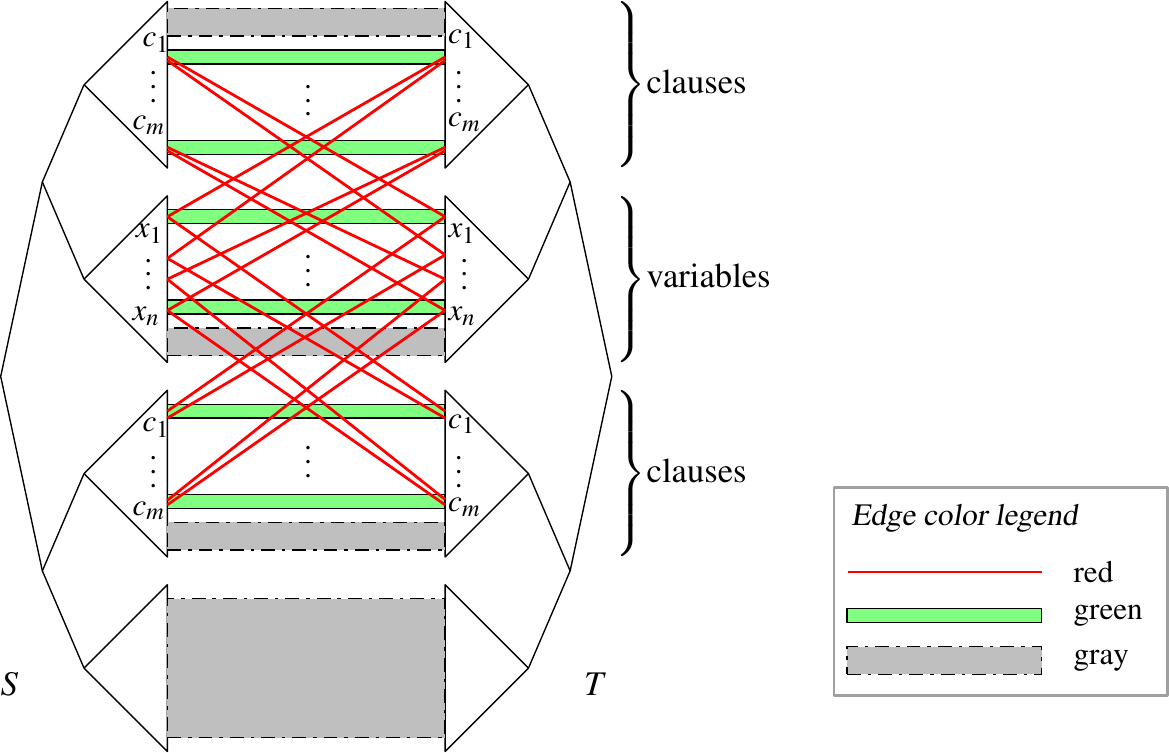}
    \caption{High-level structure of the two trees $S$ and $T$. Red
      edges connect clause and variable gadgets, green edges connect
      corresponding gadget halves, and gray edges are dummy edges
      to complete the trees.}
    \label{fig:bintrees-highlevel}
  \end{figure}

  \paragraph{Variable gadgets.} The basic structure of a variable
  gadget consists of two complete binary trees with 32 leaves each as
  shown in \figurename~\ref{fig:bintrees-var}.  Each tree has three
  highlighted subtrees of size 2 labeled $a, b, c$ and $a',b',c'$,
  respectively. From each of these subtrees there is one red
  \emph{connector} edge leaving the gadget at the top and one leaving
  it at the bottom. As long as two connector edges from the same tree
  do not cross each other, they transfer the vertical order of the
  labeled subtrees towards a clause gadget.  We define the
  configuration in \figurename~\ref{fig:bintrees-var-true} as \emph{true}
  and the configuration in \figurename~\ref{fig:bintrees-var-false} as
  \emph{false}.  If the configuration is in its \true\ state, the
  induced vertical order of the connector edges is $a < b < c$,
  otherwise the order is inverse: $c < b < a$.  It can easily be
  verified that both states have the same number of crossings.  To see
  that it is optimal observe that each pair of connector edges from
  the same subtree (for example, subtree~$a$) always crosses all 26 gray
  edges in the gadget.  Furthermore, all 24 crossings of two connector
  edges in the figure are mandatory.  Finally, the four crossings
  among the gray edges between subtrees~1 and~$2'$ and subtrees~2
  and~$1'$ are also optimal.  (Otherwise, if subtree~1 is aligned with
  subtree~$2'$, there are 12 edges from the upper subtree on the left
  to the lower subtree on the right and 10 edges from the lower
  subtree on the left to the upper subtree on the right that yield in
  total at least 120 gray--gray crossings in
  addition to the 24 red--red crossings and the 156 red--gray
  crossings as opposed to a total of 184 crossings in either
  configuration of \figurename~\ref{fig:bintrees-var}.) Note that some
  internal swaps within the subtrees 1, 2, $1'$, $2'$ are possible
  that do not affect the number of crossings; none of them, however,
  changes the order of the connector edges since in any optimal
  solution the subtrees of the four crossing gray edges must always
  stay in the center of the gadget.

  \begin{figure}[tb]
    \begin{minipage}[b]{0.61\textwidth}
      \subfloat[$x=\true$\label{fig:bintrees-var-true}]%
      {\includegraphics[page=1]{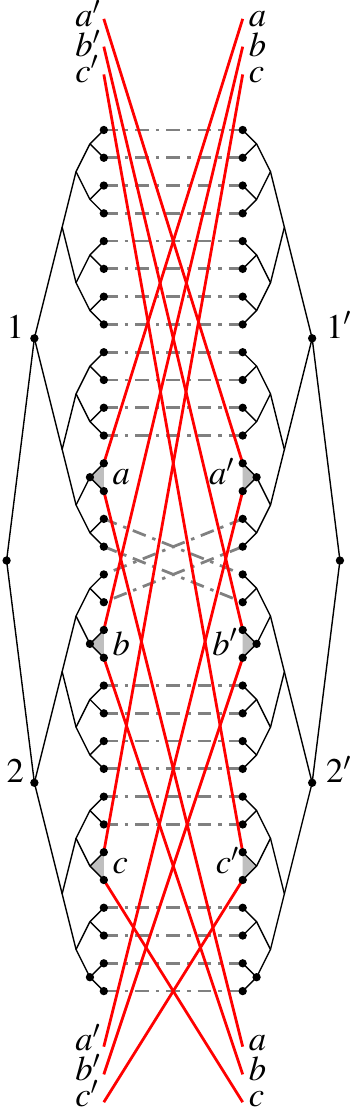}}
      \hfill \subfloat[$x=\false$\label{fig:bintrees-var-false}]%
      {\includegraphics[page=2]{variable-gadget}}

      \vspace{-2ex}
      \rule{\columnwidth}{0pt}
    \end{minipage}
    \hfill
    \begin{minipage}[b]{0.32\textwidth}
      \setcounter{subfigure}{0}
      \centering
      \subfloat[\label{fig:quaplet-gray}A single gray edge.]%
      {\quad\includegraphics{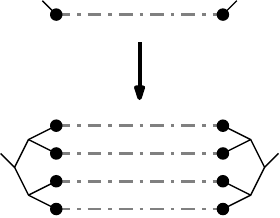}\quad\rule[-2ex]{0ex}{1ex}}\\[8ex]

      \subfloat[\label{fig:quaplet-red}Two pairs of connector edges
      for a variable used in three clauses.]%
      {\qquad\includegraphics{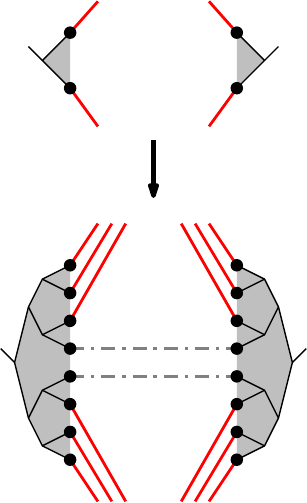}\qquad\rule[-2ex]{0ex}{1ex}}

      \vspace{-2ex}
      \rule{\columnwidth}{0pt}
    \end{minipage}

    \begin{minipage}[t]{0.61\textwidth}
      \caption{The variable gadget in its two optimal configurations
        with 184 crossings. Red edges are drawn solid, whereas
        dash-dot style is used for gray
        edges.}\label{fig:bintrees-var}
    \end{minipage}
    \hfill
    \begin{minipage}[t]{0.32\textwidth}
      \caption{Replacing each edge by four edges.}\label{fig:quaplet}
    \end{minipage}
  \end{figure}

  Note that so far the gadget in the figure is designed for a single
  appearance of the variable since the four connector-edge triplets
  are required for a single clause.  For the
  \textsc{Max2Sat} reduction, however, each variable can appear up to three
  times in different clauses.  By appending a complete binary tree
  with four leaves as in \figurename~\ref{fig:quaplet} to each leaf of the
  gadget in \figurename~\ref{fig:bintrees-var} and copying each edge
  accordingly the above arguments still hold for the enlarged trees
  with 128 leaves each.  Unused connector edges in opposite subtrees
  are linked to each other ($a$ to $a'$, $b$ to $b'$, $c$ to $c'$) as in
  \figurename~\ref{fig:quaplet-red} such that the number of crossings in
  the gadget remains balanced for both states.

  \paragraph{Clause gadgets.} For each clause $c_i = l_{i1} \lor
  l_{i2}$, where $l_{i1}$ and $l_{i2}$ denote the two literals, we
  create two clause gadgets: one in the upper clause subtrees and one
  in the lower clause subtrees (recall
  \figurename~\ref{fig:bintrees-highlevel}).  Each gadget itself consists
  of two parts: one part that uses the connectors from the first
  variable in the left tree and those from the second variable in the
  right tree and vice versa. Figure~\ref{fig:bintrees-clause} shows
  one such part of the gadget in the lower clause subtrees, where the
  connector edges lead upwards. The gadget in the upper clause subtree
  is simply a mirrored version.

  The basic structure consists of two aligned subtrees with eight
  leaves as depicted in \figurename~\ref{fig:bintrees-clause}.  Three of
  the leaves on each side serve as the missing endpoints for the
  triplets of connector edges from the corresponding variables.  Recall
  that for a positive literal with value \true\ the order of the
  connector edges is $ a < b < c$, and for a positive literal with
  value \false\ it is $c < b < a$.  (For negative literals the meaning
  of the orders is inverted.)  The two connector leaves for the edges
  labeled $a$ and $b$ are in the same four-leaf subtree, the
  connector leaf for $c$ is in the other subtree.  Three cases need to
  be distinguished.  If (1) both literals are \true, then the
  configuration in \figurename~\ref{fig:bintrees-clause-tt} is optimal with
  21 crossings. If (2) only one literal is \true, then
  \figurename~\ref{fig:bintrees-clause-tf} shows again an optimal configuration
  with 21 crossings. Here the tree on the right side swapped
  the subtrees of the root node. Finally, if (3) both literals are \false, there
  are at least 22 crossings in the gadget as shown in
  \figurename~\ref{fig:bintrees-clause-ff}.  Since this substructure is
  repeated four times for each clause we have 84 induced crossings for
  satisfied clauses and 88 induced crossings for unsatisfied clauses.

  \paragraph{Reduction.} We construct the gadgets for all variables
  and clauses and link them together as two trees $S$ and $T$, which
  are filled up with dummy leaves and edges such that they become
  complete binary trees. The general layout is as depicted in
  \figurename~\ref{fig:bintrees-highlevel}, where each dummy leaf in
  $S$ is connected to the opposite dummy leaf in $T$ such that there
  are no crossings among dummy edges.  In each of the four main
  subtrees all dummy edges are consecutive. Thus of all dummy edges
  only those in the variable subtree have crossings with exactly half
  the connector edges.

  It remains to compute the minimum number $M$ of crossings that are
  always necessary, even if all clauses are satisfied.  Then the
  \textsc{Max2Sat} instance has a solution with at least $K$
  satisfied clauses if and only if the constructed TL instance has a
  solution with at most $K' = M + 4 (|C|-K)$ crossings.  We get the
  corresponding variable assignment directly from the layout of the
  variable gadgets.

  \begin{figure}[tb]
    \begin{minipage}[b]{0.4\textwidth}
      \centering \subfloat[\label{fig:bintrees-clause-tt}$\true\lor\true$:
      21~crossings.]{\includegraphics[page=1]{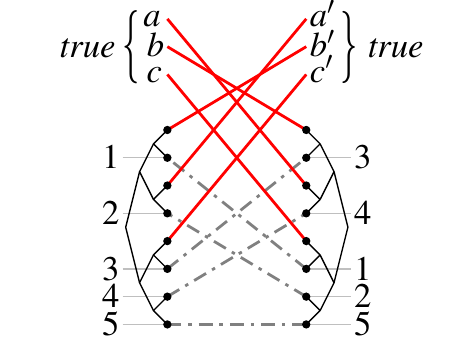}}\\ 
      \subfloat[\label{fig:bintrees-clause-tf}$\false\lor\true$:
      21~crossings.]{\includegraphics[page=2]{clause-gadget}}\\ 
      \subfloat[\label{fig:bintrees-clause-ff}$\false\lor\false$:
      22~crossings.]{\includegraphics[page=3]{clause-gadget}} 
    \end{minipage}
    \hfill
    \begin{minipage}[b]{0.6\textwidth}
      \centering
      \includegraphics{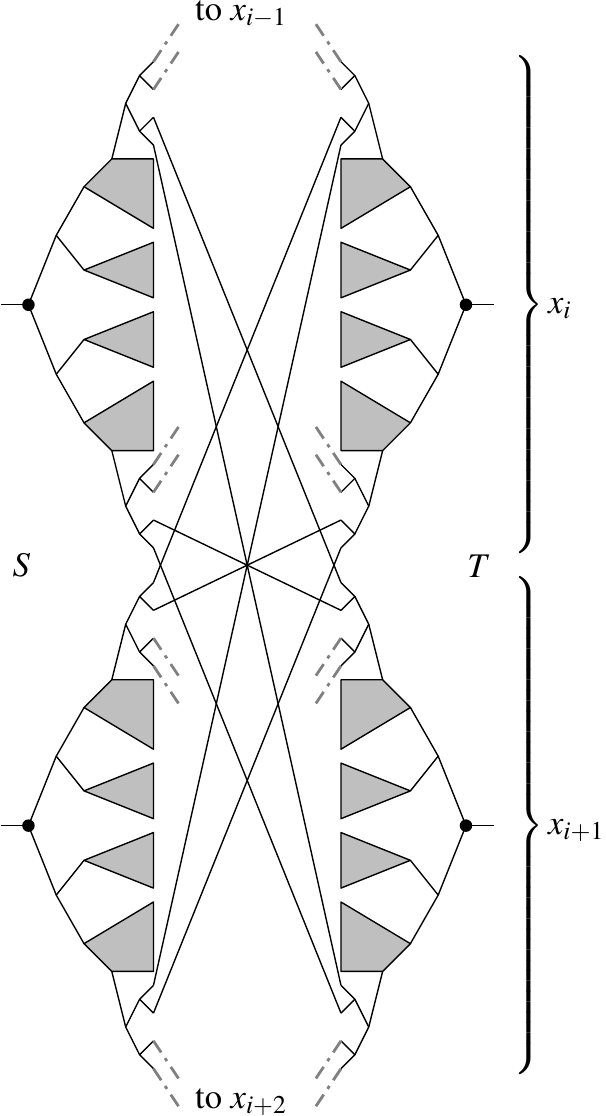}
    \end{minipage}

    \begin{minipage}[t]{0.41\textwidth}
      \caption{Gadget for the clause $c_i\,{=}\,l_{i1} \lor l_{i2}$.}
      \label{fig:bintrees-clause}
    \end{minipage}
    \hfill
    \begin{minipage}[t]{0.57\textwidth}
      \caption{Linking adjacent variable gadgets for $x_i$ and
      $x_{i+1}$.}\label{fig:link-vars}
    \end{minipage}
  \end{figure}

  The first step for computing $M$ is to fix an (arbitrary) order for
  the variable gadgets in the variable subtree. Let this order be $x_1
  < x_2 < \ldots < x_n$.  We want to achieve that any other order
  would increase the number of crossings by a number that is too large
  for it to be part of an optimum solution. We first establish
  neighbor links
  between adjacent variable gadgets. For these neighbor links we need eight of
  the 128 leaves in each half of
  each 
  variable gadget as shown in \figurename~\ref{fig:link-vars}. Since
  both subtrees below the root of $x_i$ in $S$ and both subtrees below
  the root of $x_{i+1}$ in $T$ are connected to each other, the
  minimum number of crossings of those edges is independent of the
  truth state of each gadget. The next step is to enlarge the variable
  gadgets even further by repeatedly doubling all leaves until each
  variable gadget has at least $c m^2$ gray edges for some constant
  $c$. (Note that in subtrees containing red connector edges, we do
  not duplicate any red edges but rather create new gray edges,
  similarly to \figurename~\ref{fig:quaplet-red}.) Now changing the
  variable order causes at least $8c m^2$ additional crossings since
  at least eight neighbor links would cross at least one variable
  gadget. We explain how to choose $c$ later.

  Once the order of the variables is fixed, we sort all clauses
  lexicographically (a clause with variables $x_i < x_j$ is smaller
  than a clause with variables $x_k < x_l$ if $x_i < x_k$ or if $x_i =
  x_k$ and $x_j < x_l$) and place smaller clauses towards the top of
  the clause subtrees.  Consider two clause gadgets in the same clause
  subtree.  Then, in the given clause order, there are crossings between
  their connector-edge triplets if and only if the intervals between
  their respective variables intersect in the variable order. Since
  these crossings are unavoidable for the given variable order, the
  number of connector-triplet crossings in the lexicographic order of
  the clauses is optimal. There are at most 36 crossings between the
  connector-edge triples of any pair of clause gadgets in each of the
  two clause subtrees. So for all clause pairs in both clause subtrees
  we get at most $\gamma = 2\cdot 36 \cdot m (m-1)/2$ crossings. If we
  choose the constant $c$ so that $8c m^2 > \gamma$, it never pays off
  to change the given variable order. So we can finally compute all
  necessary crossings between connector edges, dummy edges and
  intra-gadget edges which yields the number $M$.

  Since each gadget has polynomial size, the two trees and the
  number~$M$ can be computed in polynomial time. 
  It is obvious that the complete binary TL problem is in~$\mathcal{NP}$.
\end{pf}

\section{Approximation Algorithm} \label{sec:approximation}

We start with a basic observation about binary tanglegrams. As we have
noted in the introduction, TL is a purely combinatorial problem, that
is, it suffices to determine two leaf orders $\sigma$ and $\tau$ that
are compatible with the input trees $S$ and $T$, respectively. These
orders are completely determined by fixing an order of the two
subtrees of each inner node~$v \in S^\circ \cup T^\circ$, where
$S^\circ$ and $T^\circ$ denote the set of inner nodes of $S$ and
$T$. The algorithm will recursively split the two trees $S$ and $T$ at
their roots into two equally sized subinstances and determine leaf
orders of $S$ and $T$ by choosing a locally optimal order of the
subtrees below the left and right root of the current subinstance.

Let \ttree{S_0,T_0} be an input instance for complete binary TL. We assume that an
initial layout of~$S_0$ and~$T_0$ is given, that is, the subtrees of
each $v \in S_0^\circ \cup T_0^\circ$ are ordered (otherwise choose an
arbitrary initial layout). The root of a tree $T$ is denoted as
$v_T$. For a binary tree~$T$ with the two ordered subtrees~$T_1$
and~$T_2$ of~$v_T$, we use the notation $T=(T_1,T_2)$. For each
subinstance~\ttree{S,T} with $S=(S_1,S_2)$ and $T=(T_1,T_2)$, we need
to consider the four configurations $(S_1,S_2) \times (T_1,T_2)$
(initial layout), $(S_2,S_1) \times (T_1,T_2)$ (swap at $v_S$),
$(S_1,S_2) \times (T_2,T_1)$ (swap at $v_T$), and $(S_2,S_1) \times
(T_2,T_1)$ (swap at $v_S$ and $v_T$). For each configuration, we
recursively solve two subinstances and then choose the configuration
with the minimum number of crossings.

\begin{figure}[tb]
  \centering
  \includegraphics{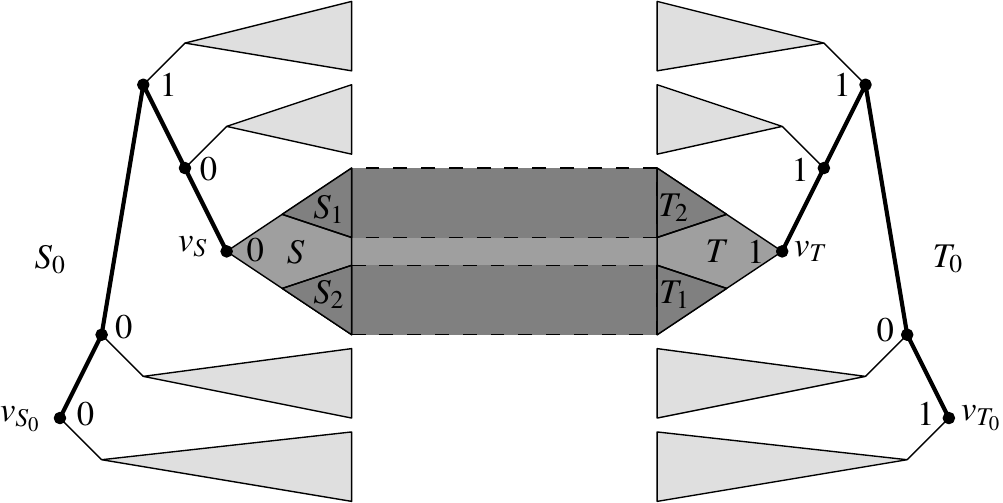}
  \caption{The context of an instance \ttree{S,T} that is split into
    the subinstances \ttree{S_1,T_2} and \ttree{S_2,T_1} since $T_1$
    and $T_2$ are swapped at $v_T$.  The swap history is indicated by
    binary swap variables along the paths to the roots $v_{S_0}$ and
    $v_{T_0}$.}
  \label{fig:subinstance}
\end{figure}

We always split the instance \ttree{S,T} into an upper and a lower
half, that is, the subinstances depend on the swap decision. If we
swap both $v_S$ and $v_T$ or none, the two subinstances
are~\ttree{S_1, T_1} and~\ttree{S_2,T_2}; if only one side is swapped,
the subinstances are~\ttree{S_1, T_2} and~\ttree{S_2, T_1}. We solve
both subinstances independently. In order to achieve the desired
approximation ratio, however, we cannot ignore the swap history of the
predecessor nodes of $v_T$ and $v_S$. This history can be regarded as
two bit strings $h_S$ and $h_T$ that represent the \emph{swap} and
\emph{no-swap} decisions made at the previous steps of the recursion.
Figure~\ref{fig:subinstance} shows an instance \ttree{S,T} and
its swap history.

The history is used to compute the number of \emph{current-level
  crossings}\index{current-level crossing} of \ttree{S,T}, that is,
the number of crossings that are caused by the swap decisions made for
the current subinstance. The number of current-level crossings and the
recursively computed numbers of crossings of the subinstances
determine which of the four configurations of the current instance is
the best one. Let $\lca(a,b)$ be the lowest common ancestor of two
nodes $a$ and $b$ of the same tree.
An important observation that is
necessary to compute the number of current-level crossings is the
following.

\begin{observation}\label{obs:lca}
  For each pair of inter-tree edges $ab$ and $cd$, $a,c \in L(S)$ and
  $b,d \in L(T)$, the swap decisions at the lowest common ancestors
  $\lca(a,c)$ and $\lca(b,d)$ completely determine whether $ab$ and
  $cd$ cross or not. Given the order of the subtrees of $\lca(a,c)$,
  swapping or not swapping the subtrees of $\lca(b,d)$ (and vice
  versa) causes or removes the crossing of~$ab$ and~$cd$.
\end{observation}

When considering the current-level crossings of a subinstance
\ttree{S,T} we know from the swap history which of the nodes on the
paths $P_S$ and $P_T$ from $v_S$ and $v_T$ to the roots~$v_{S_0}$ and
$v_{T_0}$ of the full trees, respectively, have swapped their
subtrees. Hence, for $v_S$ we can compute the current-level crossings
of all pairs of edges $ab$ and $cd$ with $a \in L(S_1)$, $c \in
L(S_2)$, and $\lca(b,d) \in P_T$; analogously, we can compute the
crossings of all pairs of edges $ab$ and~$cd$ with $b \in L(T_1)$, $d
\in L(T_2)$, and $\lca(a,c) \in P_S$. Note that if $\lca(b,d)$ or
$\lca(a,c)$ is not one of the predecessor nodes of $v_T$ or $v_S$, but
it is a node in the subtree~$T$ or~$S$, then the crossing of the edges
$ab$ and $cd$ will be considered in a subsequent step. Otherwise, our
algorithm cannot account for the crossing and we may underestimate the
number of crossings. Yet, we are able to bound this error 
later in Theorem~\ref{thm:2approx}.

\SetKwFunction{RecSplit}{RecSplit}
Algorithm~\ref{alg:2-approx} defines the recursive routine
\RecSplit that computes our tanglegram layout. It is initially
called with the parameters
\RecSplit($S_0,T_0,\varepsilon,\varepsilon$), where
$\varepsilon$ is the empty string.

\begin{algorithm}[tb]
    \DontPrintSemicolon
    \SetArgSty{}
    \KwIn{\hspace{2ex}$n$-leaf trees $S = (S_1,S_2)$ and
      $T=(T_1,T_2)$, swap histories~$h_S$ and~$h_T$} 
    \KwOut{lower bound $\cross_{ST}$ on the number of crossings
      created by the algorithm;\linebreak 
      orders $\sigma$ and $\tau$ for the leaves of $S$ and $T$, respectively}
    \eIf{$n=1$}{\KwRet{$(\cross_{ST}, \sigma, \tau) = (0,v_S,v_T)$}\;}
    {
      $\cross_{ST} = \infty$\;
      \ForEach{$(\swapl,\swapr)
        \in \{0, 1\}^2$}{
        \emph{loop through all four cases to swap subtrees of $S$ and $T$}\;
        $\cl \leftarrow$ current level crossings induced by $(\swapl,\swapr)$\;
        $(\cross_1,\sigma_{1+\swapl},\tau_{1+\swapr}) \leftarrow
        \RecSplit(S_{1+\swapl},T_{1+\swapr},(h_S,\swapl),(h_T,\swapr))$\;
        $(\cross_2,\sigma_{2-\swapl},\tau_{2-\swapr}) \leftarrow
        \RecSplit(S_{2-\swapl},T_{2-\swapr},(h_S,\swapl),(h_T,\swapr))$\;
        \If{$\cl + \cross_1 + \cross_2 < \cross_{ST}$}{
          $\cross_{ST} \leftarrow \cl + \cross_1 + \cross_2$\;
          \uIf{$\swapl=0$}{$\sigma \leftarrow (\sigma_1,\sigma_2)$\;}
          \lElse{$\sigma \leftarrow (\sigma_2,\sigma_1)$\;}
          \uIf{$\swapr=0$}{$\tau \leftarrow (\tau_1,\tau_2)$\;}
          \lElse{$\tau \leftarrow (\tau_2,\tau_1)$\;}
        }
      }
      \KwRet{$(\cross_{ST}, \sigma, \tau)$}\;
    }
    \caption{\protect\RecSplit($S,T,h_S,h_T$)}\label{alg:2-approx}
\end{algorithm}

In order to quickly calculate the number of current-level crossings we
use a preprocessing step. To that end, we compute two tables $C^=$ and
$C^\times$ of size $O(n^2)$. For each pair $(v,w)$ of inner nodes in
$S^\circ \times T^\circ$, the entry $C^=[v,w]$ stores the number of
crossings of edge pairs~$ab$ and $cd$ with $\lca(a,c)=v$ and
$\lca(b,d)=w$ if either both or none of $v$ and $w$ swap their
subtrees. An entry $C^\times[v,w]$ stores the analogous number of
crossings if only one of~$v$ and~$w$ swap their subtrees.

\begin{lemma}\label{lem:cross-tables}
  The tables $C^=$ and $C^\times$ can be computed in $O(n^2)$ time.
\end{lemma}

\begin{pf}
  We initialize all entries as~0 and preprocess $S_0$ and $T_0$ in
  linear time to support lowest-common-ancestor queries in $O(1)$
  time~\cite{gt-ltasc-83}. Then we determine for each pair of
  inter-tree edges their lowest common ancestors in $S_0$ and $T_0$
  and increment the corresponding table entry depending on which two
  configurations yield the crossing. This takes $O(n^2)$ time for all
  edge pairs.
\end{pf}

Once we have computed~$C^=$ and~$C^\times$, we can determine the number
of current-level crossings for any subinstance \ttree{S,T} in $O(\log n)$
time by summing up the appropriate table entries depending on the swap
history along the paths~$P_T$ and~$P_S$, which are of length $O(\log n)$.

The running time Algorithm~\ref{alg:2-approx} satisfies the recurrence
$T(n) \leq 
8 T(n/2) + O(\log n)$, which solves to $T(n)=O(n^3)$ by the master
method~\cite{clrs-ita-01}.  We now prove that the algorithm yields a
2-approximation.

\begin{theorem}\label{thm:2approx}
  Given a complete binary TL instance \ttree{S_0,T_0} with
  $n$ leaves in each tree, Algorithm~\ref{alg:2-approx} computes in
  $O(n^3)$ time a drawing of \ttree{S_0,T_0} that has at most twice as
  many crossings as an optimal drawing.
\end{theorem}

\begin{pf}
  Fix any drawing~$\delta$ of \ttree{S_0,T_0}. 
  Algorithm~\ref{alg:2-approx} tries,
  for each subinstance \ttree{S,T} of \ttree{S_0,T_0}, all four
  possible configurations of $S=(S_1, S_2)$ and $T=(T_1, T_2)$---among
  them the configuration in $\delta$. Assume that the configuration
  in~$\delta$ is \ttree{(S_1,S_2),(T_1,T_2)}. We determine an upper bound
  on the number of crossings that the algorithm fails to count for the
  drawing~$\delta$. In each of the trees $S_0$ and $T_0$ we
  distinguish four different areas for the endpoints of the edges:
  above $S_1$, in $S_1$, in $S_2$, below $S_2$ and similarly above
  $T_1$, in $T_1$, in $T_2$, below $T_2$.  We number these regions
  from~0 to~3, see \figurename~\ref{fig:edge-types}. This allows
  us to classify the edges into 16 groups (two of which, 0--0 and
  3--3, are not relevant). We denote the number of \emph{$i$--$j$
    edges}, that is, edges from area~$i$ to area~$j$, by $n_{ij}(S,T)$
  (for $i,j \in \{0,1,2,3\}$). Figure~\ref{fig:n1*types} shows
  the four groups of $i$--$j$ edges for $i=1$.

  \begin{figure}[tb]
    \centering
    \subfloat[Edges incident to $L(S_1)$ are separated into four
    groups by the area of their second endpoint.\label{fig:n1*types}]%
    {\quad\includegraphics{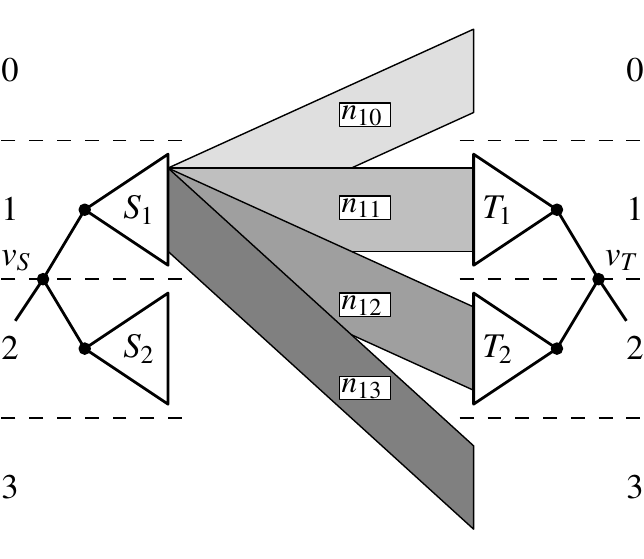}\quad}
    \hfill
    \subfloat[All edge groups that cross the $n_{12}$ 1--2 edges between
    $L(S_1)$ and $L(T_2)$.\label{fig:n12-crossings}]%
    {\quad\includegraphics{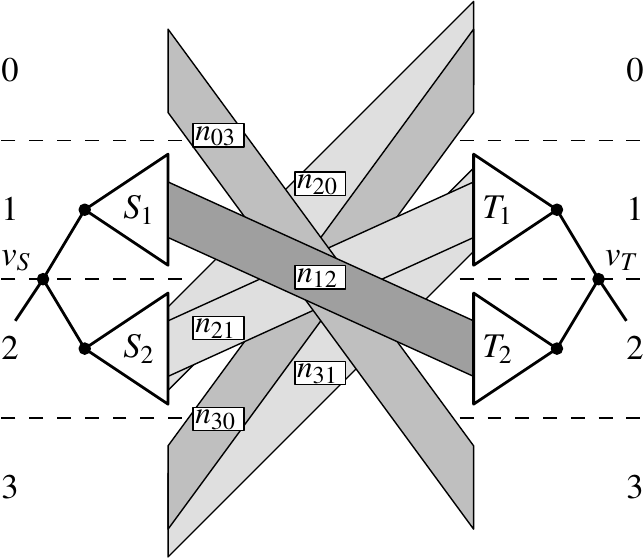}\quad}
    \caption{Areas of the endpoints and types of edges incident to
      $L(S)$ and $L(T)$. Cardinalities $n_{ij}(S,T)$ are abbreviated
      as $n_{ij}$.}
    \label{fig:edge-types}
  \end{figure}

  The only crossings that the algorithm does not take into account are
  crossings between edges whose lowest common ancestors lie in parts
  of $S_0$ and $T_0$ that are split apart into different branches of
  the recursion. For the subinstance \ttree{S,T}, which is split into
  \ttree{S_1,T_1} and \ttree{S_2,T_2}, this means that for all
  $n_{12}(S,T)$ edges that run between $S_1$ and $T_2$, we fail to
  consider all crossings between pairs of two such edges. Similarly,
  we do not consider any pair of the $n_{21}(S,T)$ edges between $S_2$
  and $T_1$.

  Let's return to the drawing $\delta$ and consider the set
  $\mathcal{I}$ of subinstances that correspond to~$\delta$, that is,
  all pairs of opposing subtrees in $\delta$. For each subinstance
  $\ttree{S,T} \in \mathcal{I}$ we do not account for crossings of
  pairs of 1--2 edges and pairs of 2--1 edges since these edges run
  between two subinstances that are solved independently. In the worst
  case all these edge pairs cross and the algorithm misses
  $\binom{n_{12}(S,T)}{2} + \binom{n_{21}(S,T)}{2}$ crossings. Let
  $c_\delta$ be the number of crossings of $\delta$ counted by the
  algorithm, and let $|\delta|$ be the actual number of crossings
  of~$\delta$.  Clearly, we have $c_\delta \le |\delta|$. We can
  bound $|\delta|$ from above by
  \begin{equation}
    \label{eq:approx}
    |\delta| \le c_\delta + \sum_{\ttree{S,T} \in
      \mathcal{I}} \left[ \binom{n_{12}(S,T)}{2} +
      \binom{n_{21}(S,T)}{2} \right] \le c_\delta + \sum_{\ttree{S,T} \in
      \mathcal{I}} \frac{n_{12}^2(S,T) + n_{21}^2(S,T)}{2}.
  \end{equation}

  We now show that $\sum_{\ttree{S,T} \in \mathcal{I}} (n_{12}^2(S,T)
  + n_{21}^2(S,T)) \le 2 c_\delta$.  For the sake of convenience,
  we abbreviate~$n_{ij}(S,T)$ by~$n_{ij}$ in the following. We will
  bound $n_{12}^2$ by the number of crossings of the 1--2 edges in
  $\delta$ that are counted by the algorithm. This number is at
  least
  \begin{equation}
    \label{eq:c12}
    c_{12} = n_{12} \cdot (n_{03} + n_{20} + n_{21} + n_{30} +
    n_{31})
  \end{equation}
  as can be seen in \figurename~\ref{fig:n12-crossings}. All
  these crossings are current-level crossings at this or some earlier
  point in the algorithm. Since our (sub)trees are complete and thus
  $S_1$ and $T_1$ have the same number of leaves, we obtain
  \begin{equation}\label{eq:Balance_equality_1}
    n_{10}+n_{12}+n_{13} = n_{01}+n_{21}+n_{31}.
  \end{equation}
  Furthermore, we have the following equality for the edges from
  areas 0 on both sides
  \begin{equation}
    \label{eq:area0}
    n_{01} + n_{02} + n_{03} = n_{10} + n_{20} + n_{30}.
  \end{equation}
  From (\ref{eq:Balance_equality_1}) we obtain $n_{12} \le
  n_{01}-n_{10}+n_{21}+n_{31}$ and from (\ref{eq:area0}) we
  obtain $n_{01}-n_{10} \leq n_{20}+n_{30}$.  Hence, we have $n_{12}
  \le n_{20} + n_{30} + n_{21} + n_{31}$. With (\ref{eq:c12})
  this yields
  \begin{equation}
    \label{eq:case1}
    n_{12}^2 \le n_{12} \cdot (n_{20} + n_{30} + n_{21} + n_{31}) \le c_{12},
  \end{equation}
  that is, $n_{12}^2$ is bounded by the number of crossings that
  involve a 1--2 edge in $\delta$ and that are counted by the
  algorithm.  Analogously, we obtain
  \begin{equation}
    \label{eq:case1-2}
    n_{21}^2 \le n_{21} \cdot (n_{02} + n_{03} + n_{12} + n_{13}) \le c_{21},
  \end{equation}
  that is, $n_{21}^2$ is bounded by the number of crossings counted by
  the algorithm that involve a~2--1 edge in $\delta$.

  So from~(\ref{eq:case1}) and~(\ref{eq:case1-2}) we have
  $n_{12}^2 \le c_{12}$ and $n_{21}^2 \le c_{21}$. Applying this
  argument to all subinstances $\ttree{S,T} \in \mathcal{I}$ we get
  \begin{equation}
    \label{eq:le2opt}
    \sum_{\ttree{S,T} \in \mathcal{I}} (n_{12}^2(S,T) + n_{21}^2(S,T))
    \le \sum_{\ttree{S,T} \in \mathcal{I}} c_{12}(S,T) +
    \sum_{\ttree{S,T} \in \mathcal{I}} c_{21}(S,T) \le 2 \cdot
    c_\delta.
  \end{equation}
  The fact that $\sum_{\ttree{S,T} \in \mathcal{I}} c_{12}(S,T) \le
  c_\delta$ holds is due to each edge crossing~$\delta$ appearing in at
  most one term~$c_{12}(S,T)$.
  This can be seen as follows.
  Let $ab$ be a 1--2 edge in the
  subinstance \ttree{S,T}. Then in all parent instances of the
  recursion, $ab$ was still a 1--1 edge or a 2--2 edge; such edges do
  not appear in any previous $c_{12}$-term. In a subsequent instance
  \ttree{S',T'} below \ttree{S,T} in the recursion the edge $ab$ might
  in fact reappear, for example as a 0--3 edge. At that point, however,
  it is considered as an edge that crosses one of the 1--2 edges of
  \ttree{S',T'}, say~$cd$. But then $cd$ was considered as a 1--1 or
  2--2 edge in all previous instances.  Hence, the crossing between $ab$
  and $cd$ does not appear in any other $c_{12}$-term. Analogous
  reasoning yields~$\sum_{\ttree{S,T} \in \mathcal{I}} c_{21}(S,T) \le
  c_\delta$

  Plugging~(\ref{eq:le2opt}) into~(\ref{eq:approx}) yields
  $|\delta| \le 2 c_\delta$. Now let $A^\star$ be the solution
  computed by Algorithm~\ref{alg:2-approx} and let~$S^\star$ be an optimal
  solution.  We denote their actual numbers of crossings
  by~$|A^\star|$ and~$|S^\star|$, respectively. By~$c_{A^\star}$
  and~$c_{S^\star}$ we denote the number of crossings counted by our 
  algorithm for the drawings~$A^\star$ and~$S^\star$, respectively.
  Since $|\delta| \le 2 c_\delta$ for any drawing~$\delta$ we get
  \[|A^\star| \le 2 c_{A^\star} \le 2 c_{S^\star} \le 2 |S^\star|,\]
  that is, the algorithm is indeed a factor-2 approximation.
\end{pf}

\begin{figure}[tb]
  \centering
  \includegraphics{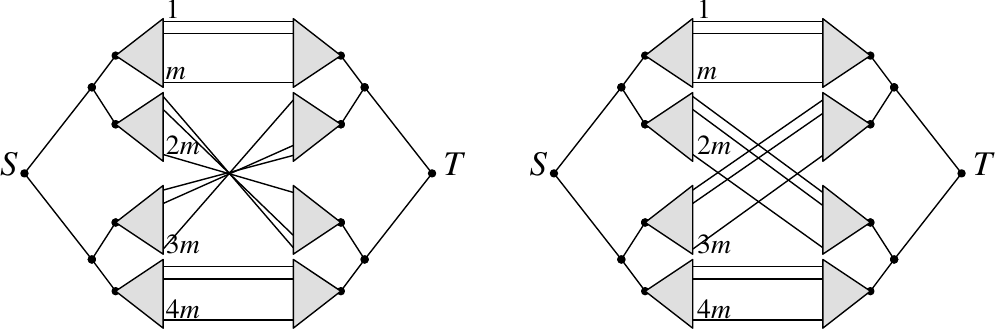}
  \caption{Example of a tanglegram for which our algorithm may output
    a drawing (left) that has roughly twice as many crossings as the
    optimal drawing (right).}
  \label{fig:Factor2Tight}
\end{figure}

We note that the approximation factor of~2 is tight:
let $n=4m$, let $S$ have leaves ordered $1,\dots,4m$, and let $T$ have
leaves ordered $1,\ldots,m,3m,\ldots,2m+1,$ $m+1,\ldots,2m,3m+1,\ldots,4m$
(see \figurename~\ref{fig:Factor2Tight}).  Then our algorithm may
construct a drawing with $m^2 + 2\binom{m}{2} = 2m^2 -m $ crossings,
while the optimal drawing has only $m^2$ crossings.

\paragraph{Non-complete binary trees.} Algorithm~\ref{alg:2-approx} can
also be applied to non-complete tanglegrams with minor
modifications. The only essential difference is that during the
algorithm we can encounter the situation that a single leaf $v$ of one
tree is paired with a larger subtree $T'$ of the other tree. In that case
we continue the recursion for those subtrees of $T'$ that contain an
edge to $v$ in order to find their locally optimal swap decisions. For
non-complete tanglegrams, however,
the approximation factor does not hold any more.  N\"ollenburg et
al.~\cite{nvwh-dbtee-09} have evaluated several heuristics for binary TL,
among them the modified version of Algorithm~\ref{alg:2-approx}.

\paragraph{Generalization to $d$-ary trees.}
The algorithm can be generalized to complete $d$-ary trees. The
recurrence relation of the running time changes to $T(n) \leq d \cdot
(d!)^2 \cdot T(n/d) + O(\log n)$ since we need to consider all~$d!$ subtree
orderings of both trees, each triggering $d$ subinstances of size
$n/d$. This resolves to $T(n)=O(n^{1 + 2 \log_d (d!)})$. For $d \ge 3$
the running time is upper-bounded by $O(n^{2d-1.7})$. At the same time
the approximation factor increases to $1+\binom{d}{2}$. This is
because for any pair $(i,j)$ with $1\le i < j \le d$ the algorithm
fails to account for potential crossings between the trees $S_i$ and
$T_j$ as well as between $S_j$ and $T_i$. This number can be bounded
for each of the $\binom{d}{2}$ pairs by the number of crossings in the
optimal solution using our arguments for binary trees.

\paragraph{Maximization version.}
Instead of the original TL problem, which minimizes the number of
pairs of edges that cross each other, we now consider the dual problem
TL$^\star$ of maximizing the number of pairs of edges that do not
cross.  The sets of optimal solutions for the two problems are the
same, but from the perspective of approximation the problems differ
a lot, at least in the binary case: in contrast to binary TL, which is
hard to approximate as we have shown in Theorem~\ref{thm:hard}, binary
TL$^\star$ has a constant-factor approximation algorithm.  We show
this by reducing binary
TL$^\star$ to a constrained version of the \textsc{MaxCut} problem,
which can be solved approximately with the semidefinite programming
(SDP) rounding algorithm of Goemans and Williamson~\cite{gw-iaamc-95}.
Their algorithm runs in polynomial time; solving the underlying SDP
relaxation of the problem is the most time-consuming step.
Still, SDP relaxations of \textsc{MaxCut} instances of up to 7000
variables can be solved in practice~\cite{bm-pgasm-01}.

\begin{theorem}
  There exists a polynomial-time factor-$0.878$ approximation
  algorithm for binary TL$^\star$.
\end{theorem}

\begin{pf}
  Let \ttree{S,T} be an instance of binary TL$^\star$.  Fix any
  initial drawing of \ttree{S,T}.  As before, we associate a decision
  variable with each inner node of the two trees.  The variable
  decides whether we do or do not swap the children at the
  corresponding node.  We model this situation by a weighted graph
  $G=(V,E)$; a swap decision corresponds to deciding to which side of
  a cut the corresponding vertex is assigned.  More precisely, for
  each inner node~$u$ of \ttree{S,T}, the graph~$G$ contains two
  vertices~$u$ and~$u'$. We will also impose a constraint that
  $u$ and $u'$ must be separated by a cut we are looking for.
  As we will indicate later, we can use the algorithm of Goemans and
  Williamson~\cite{gw-iaamc-95} to find large cuts among those
  separating all pairs of type~$(u,u')$. 

  For each pair~$ab$ and~$cd$ of inter-tree
  edges with $a,c \in L(S)$ and $b,d \in L(T)$, the graph~$G$ contains
  a weighted edge that we construct as follows.  Let $v = \lca(a,c)$
  and $w = \lca(b,d)$ be the lowest common ancestors of the edge pair.
  If~$ab$ and~$cd$ cross in the initial drawing, we add the edge~$vw$
  with weight~1 to~$G$.  If the edge is already present, we increase
  its weight by one.  If the two edges do not cross in the initial
  drawing, then we analogously add the edge~$vw'$ to~$G$ or increase
  its weight by one.

  Consider a cut in~$G$ that for each inner node~$u$ of \ttree{S,T}
  separates~$u$ and~$u'$.  We claim that any such cut encodes a
  drawing of \ttree{S,T}.  To see this, let $(F, N=(V \setminus F))$
  be such a cut.  Starting from the initial drawing we construct a new
  drawing as follows.  Let~$u$ be an inner node of \ttree{S,T}.  If $u
  \in F$ and $u' \in N$, we swap the children of the inner node~$u$ of
  the current drawing.  If $u \in N$ and $u' \in F$, we do nothing.
  (Note that exchanging the roles of the sets~$F$ and~$N$ yields the
  mirrored drawing with the same number of crossings.)

  For a moment, think of $G$ as of a multigraph that is obtained
  by replacing each edge of weight~$k$ by~$k$ edges of weight one.
  Let us argue that the above described procedure to decode
  drawings from cuts has the property that in the resulting drawing of
  \ttree{S,T}, pairs of inter-tree edges that do not cross correspond
  one-to-one to edges in~$G$ that are cut by~$(F,N)$. 
  Consider first the cut corresponding to the initial drawing,
  namely the cut with $u \in N$ for each inner node~$u$ of \ttree{S,T}
  and observe that the claim holds for this cut. Now consider a single
  swap operation at an inner node~$u$ of \ttree{S,T} and the
  corresponding change in the cut.
  Note that it changes the ``cut status'' of exactly those pairs of edges
  that have $u$ as the lowest common ancestor of two of their
  endpoints; at the same time
  it also changes the cut status of exactly the edges in~$G$ corresponding
  to these pairs of edges in the drawing. Since any cut in~$G$ may be
  reached by a finite sequence of such swap operations from the initial one,
  the property holds for any cut.
  Therefore, the number of pairs of non-crossing inter-tree
  edges in the obtained drawing equals the total weight of the cut
  (in the original, weighted version of $G$).

  The resulting optimization problem is the \textsc{MaxResCut}
  problem, that is, \textsc{MaxCut} with additional constraints
  forcing certain pairs of vertices to be separated by the cut.
  Goemans and Williamson~\cite{gw-iaamc-95}, when describing their
  famous algorithm for the \textsc{MaxCut}
  problem, observed that adding constraints to separate certain pairs
  of vertices does not make the problem harder to approximate. It is sufficient
  to encode these constraints as additional linear constraints in the
  SDP relaxation and to observe that random hyperplanes used to
  separate vertices always separate such constrained pairs.

  We use their SDP rounding algorithm for \textsc{MaxResCut} to compute a
  $0.878$-approximation of the largest cut in~$G$.  This cut
  determines which of the subtrees in the initial drawing must be
  swapped to obtain a drawing that is a $0.878$-approximation to binary
  TL$^\star$.
\end{pf}

Note that our proof also works in a slightly more general case, namely
for pairs of (not necessarily binary) trees where for each inner node
the only choice for
arranging the children is between a given permutation and the reverse
permutation obtained by swapping the whole block of children.

\section{Fixed-Parameter Tractability} \label{sec:fpt}

We consider the following parameterized variant of the complete binary
TL problem. 
Given a complete binary TL instance $\ttree{S,T}$ and a non-negative
integer $k$, decide whether there exists a layout of $S$ and $T$ with
at most $k$ induced inter-tree edge crossings.  Our algorithm makes
use of the same technique to count current-level crossings as the
2-approximation algorithm.  Hence, we precompute the crossing
tables~$C^=$ and~$C^\times$ in $O(n^2)$ time as before, see 
Lemma~\ref{lem:cross-tables}. The algorithm traverses the inner
nodes of $S$ in breadth-first order. It starts at the root of~$S$ and
its corresponding node in~$T$ (in this case the root of $T$), branches
into all four possible subtree configurations (at the root it actually
suffices to consider two of them), and subtracts from $k$ the number
of current-level crossings in each branch. Then we proceed recursively with
the next node $v$ in $S$, its corresponding opposite node $w$ in $T$,
and the reduced parameter $k'$ of allowed crossings. In each node of
the search tree we count the current-level crossings for each of the
subtree orders of $v$ and $w$ by summing up in linear time the
appropriate entries in $C^=$ and $C^\times$ for $v$ (or $w$) and all
of the $O(n)$ subtree orders that are already fixed in $T$ (or
$S$). Once we reach a leaf of the search tree we know the exact number
of crossings since each pair of edges $ab$ and $cd$ is counted as soon
as the subtree orders of both $\lca(a,c)$ and $\lca(b,d)$ are
fixed. Obviously, we stop following a branch of the search tree when
the parameter value drops below~0.

For the search tree to have bounded height, we need to ensure that
whenever we move to the next subinstance, the parameter value
decreases at least by one.  At first sight this seems problematic: if
a subinstance does not incur any current-level crossings, the
parameter will not drop.  The following key lemma---which does not
hold for non-complete binary trees---shows that there is a way out. It says
that if there is an order of the subtrees in a subinstance that does
not incur any current-level crossings, then we can ignore the other
three subtree orders and do not have to branch.

\begin{lemma}\label{lem:zero-crossing}
  Let $\ttree{S,T}$ be a complete binary TL instance, and let~$v_S$ be
  a node of~$S$ and~$v_T$ a node of~$T$ such that~$v_S$ and~$v_T$ have
  the same distance to their respective root.  Further, let
  $(S_1,S_2)$ be the subtrees incident to~$v_S$ and let $(T_1,T_2)$ be
  the subtrees incident to~$v_T$. If the subinstance
  \ttree{(S_1,S_2),(T_1,T_2)} does not incur any current-level
  crossings, then each of the subinstances
  \ttree{(S_1,S_2),(T_2,T_1)}, \ttree{(S_2,S_1),(T_1,T_2)}, and
  \ttree{(S_2,S_1),(T_2,T_1)} has at least as many crossings as the
  instance 
  \ttree{(S_1,S_2),(T_1,T_2)}, for any fixed ordering of the leaves
  of~$S_1$, $S_2$, $T_1$ and~$T_2$.
\end{lemma}

\begin{pf}
  If the subinstance \ttree{(S_1,S_2),(T_1,T_2)} does not incur any
  current-level crossings, this excludes certain types of edges. We
  categorize the inter-tree edges originating from the four subtrees
  according to their destinations as before, and use the notation
  $n_{ij}$ for the number of edges between area~$i$ on the left and
  area~$j$ on the right---see \figurename~\ref{sfg:S12-T12}. First of
  all, there are no edges between~$S_1$ and~$T_2$ or between~$S_2$
  and~$T_1$.  We consider only the first case, that is, $n_{12}=0$; the 
  second case $n_{21}=0$ is symmetric. In both cases, we have $n_{13}
  = n_{31} = n_{20} = n_{02} = 0$.  Since we consider complete binary
  trees, we obtain the three equalities $n_{10} = n_{01} + n_{21}$, 
  $n_{32} = n_{23} + n_{21}$, and $n_{01} + n_{11} = n_{23} + n_{22}$.

  \begin{figure}[tb]
    \centering
    \subfloat[\ttree{(S_1,S_2),(T_1,T_2)}\label{sfg:S12-T12}]%
    {\includegraphics{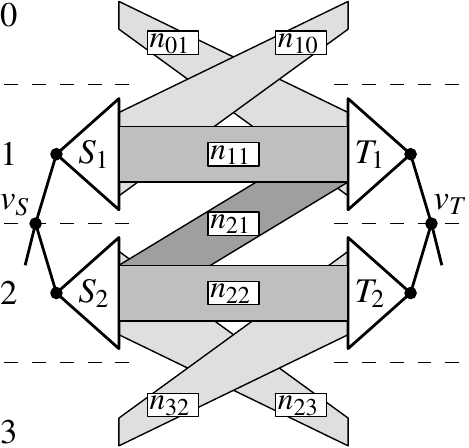}}
    \hfill
    \subfloat[\ttree{(S_2,S_1),(T_2,T_1)}\label{sfg:S21-T21}]%
    {\includegraphics{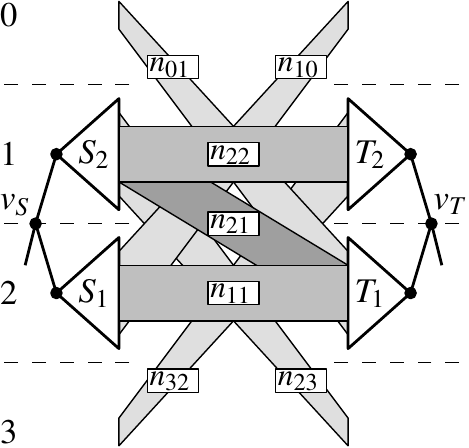}}
    \hfill
    \subfloat[\ttree{(S_1,S_2),(T_2,T_1)}\label{sfg:S12-T21}]%
    {\includegraphics{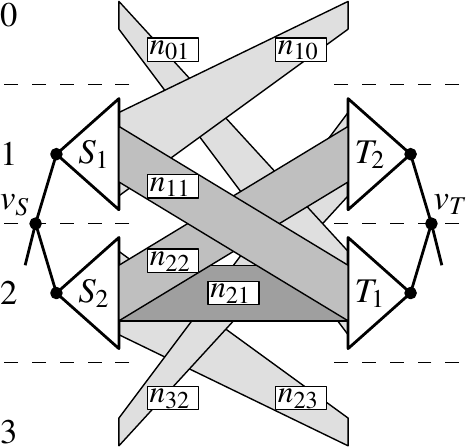}}
    \caption{Edge types and crossings of the instance
      $\ttree{S,T}$. Only non-empty classes of edge types are shown.}
    \label{fig:zero-crossings}
  \end{figure}

  We fix an ordering~$\sigma$ of the leaves of the four subtrees
  $S_1,S_2,T_1,$ and $T_2$.  We first compare the number of crossings
  in the subinstance \ttree{(S_1,S_2),(T_1,T_2)} with the number of
  crossings in the subinstance \ttree{(S_2,S_1),(T_2,T_1)}, see
  Figures~\ref{sfg:S12-T12} and~\ref{sfg:S21-T21}.  The subinstance
  \ttree{(S_1,S_2),(T_1,T_2)} can have at most $n_{21}
  (n_{11}+n_{22})$ crossings that do not occur in
  \ttree{(S_2,S_1),(T_2,T_1)}.  However, \ttree{(S_2,S_1),(T_2,T_1)}
  has at least $n_{10} (n_{23} + n_{21} + n_{22}) + n_{23} n_{11} +
  n_{32} (n_{01} + n_{21} + n_{11}) + n_{01} n_{22}$ crossings that do
  not appear in \ttree{(S_1,S_2),(T_1,T_2)}.  Plugging in the above
  equalities for $n_{10}$ and $n_{32}$, we get $(n_{01} + n_{21})
  (n_{23} + n_{21} + n_{22}) + n_{23} n_{11} + (n_{23} + n_{21})
  (n_{01} + n_{21} + n_{11}) + n_{01} n_{22} \ge n_{21}
  (n_{11}+n_{22})$.  Thus, the subinstance \ttree{(S_2,S_1),(T_2,T_1)}
  has at least as many crossings with respect to the fixed leaf
  order~$\sigma$ as \ttree{(S_1,S_2),(T_1,T_2)} has.

  Next, we compare the number of crossings in the subinstance
  \ttree{(S_1,S_2),(T_1,T_2)} with the number of crossings in the
  subinstance \ttree{(S_1,S_2),(T_2,T_1)}, see
  Figures~\ref{sfg:S12-T12} and~\ref{sfg:S12-T21}.  Now the number of
  additional crossings of \ttree{(S_1,S_2),(T_1,T_2)} is at most
  $n_{21} n_{22}$, and the subinstance \ttree{(S_1,S_2),(T_2,T_1)}
  introduces at least $(n_{01} + n_{11})(n_{32} + n_{22}) + n_{32}
  n_{21}$ additional crossings. With the equality $n_{01} + n_{11} =
  n_{23} + n_{22}$ and the inequality $n_{32} + n_{22} \ge n_{21}$ we
  get $(n_{01} + n_{11}) (n_{32} + n_{22}) + n_{32} n_{21} \ge
  (n_{23}+n_{22}+n_{32}) n_{21} \ge n_{22}
  n_{21}$.  Thus, the subinstance \ttree{(S_1,S_2),(T_2,T_1)} has at
  least as many crossings with respect to~$\sigma$ as
  \ttree{(S_1,S_2),(T_1,T_2)} has.

  By symmetry, the same holds for the last case
  \ttree{(S_2,S_1),(T_1,T_2)}, which incurs at least as many crossings
  as $n_{11}n_{21}$, the number of crossings that can be present in
  \ttree{(S_1,S_2),(T_1,T_2)} but not in \ttree{(S_2,S_1),(T_1,T_2)}.
\end{pf}

Counting the current-level crossings takes $O(n)$ time for each node
that fixes its subtree order. If an order does not incur any
current-level crossings we might need to fix in total up to $O(n)$
subtree orders and count the incurred crossings until we reach a new
node of the search tree. Thus we spend $O(n^2)$ time for each of the
$O(4^k)$ search-tree nodes. Including the preprocessing this yields a
total running time of $O(n^2 + 4^k n^2)$. If the algorithm reaches a
leaf of the search tree it has fixed all subtree orders in~$S$ and~$T$
and thus found a layout of the input instance that has at most~$k$
inter-tree edge crossings. If the search stops without reaching a leaf
there is no layout of \ttree{S,T} with at most~$k$ inter-tree edge
crossings.

\begin{theorem}
  Given a complete binary TL instance \ttree{S,T} with~$n$ leaves in
  each tree and an integer~$k$, in~$O(4^k n^2)$ time we can either
  determine a layout of \ttree{S,T} with at most~$k$ inter-tree edge
  crossings or report that no such layout exists.
\end{theorem}

\begin{figure}[tb]
  \centering
  \subfloat[\label{fig:zero-crossing-opt}]{\includegraphics{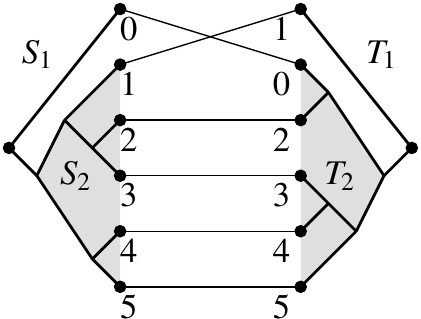}}
  \hfill
  \subfloat[\label{fig:zero-crossing-bad}]{\includegraphics{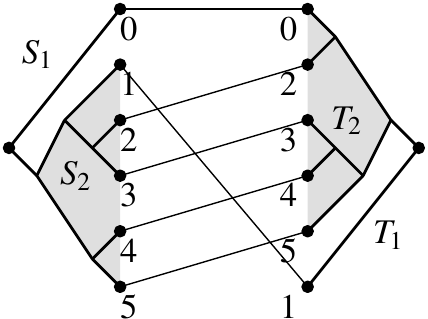}}
  \hfill
  \subfloat[\label{fig:zero-crossing-subopt}]{\includegraphics{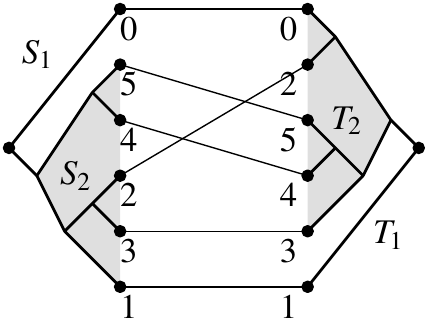}}
  \caption{Example of a binary TL instance with an optimal layout that has
    one crossing (a). The same order of the leaves in the subtrees
    $S_2$ and $T_2$ yields four crossings for a configuration without
    current-level crossings (b). The best layout that avoids the
    current-level crossing still has two crossings (c).}
  \label{fig:zero-crossing-general}
\end{figure}

Finally, the fact that Lemma~\ref{lem:zero-crossing} relies on the
completeness of the two trees is illustrated in
\figurename~\ref{fig:zero-crossing-general}. Here we have an
example of an instance whose optimal layout requires a current-level
crossing (\figurename~\ref{fig:zero-crossing-opt}). At the same
time, the configuration \ttree{(S_1,S_2),(T_2,T_1)} has no
current-level crossing. According to Lemma~\ref{lem:zero-crossing} the
leaf order of the optimal layout copied into the layout without
current-level crossings would produce at most as many crossings as in
the other layout. Figure~\ref{fig:zero-crossing-bad} shows that
this is not true in our example. The best solution of the
configuration \ttree{(S_1,S_2),(T_2,T_1)} still has two crossings and
is not optimal
(\figurename~\ref{fig:zero-crossing-subopt}). Hence, we do have
to consider \emph{all} subtree orders even if one of them incurs no
current-level crossings. This means that we cannot bound the size of
the search tree in terms of the parameter~$k$ as we have done for
complete binary trees.

\section{Open Problems} \label{sec:conclusions}

We have shown that one cannot expect to find a constant-factor
approximation for binary TL.  Would it help if \emph{one} of
the two given trees was complete? We have given a factor-2
approximation for complete binary TL.  It is natural to ask whether we
can do better.

An alternative optimization goal is to remove a minimum number of
inter-tree edges in order to obtain a planar tanglegram.

\phantomsection
\addcontentsline{toc}{section}{\numberline{}Acknowledgments}
\section*{Acknowledgments}

We thank Danny Holten and Jack van Wijk for
introducing us to this exciting problem and David Bryant for pointing us
to the work of Roderic Page on host and parasite trees.

\phantomsection
\addcontentsline{toc}{section}{\numberline{}\refname}
\bibliographystyle{alpha}
\bibliography{abbrv,twotrees}

\end{document}